\documentstyle[12pt]{article}
\newcommand{\be}{\begin{equation}}
\newcommand{\ee}{\end{equation}}
\begin{document}
\begin{titlepage}
\begin{flushright}
{\bf HU-SEFT R 1996-04}\\
\end{flushright}
\vspace{1cm}
\begin{center}
{\Large \bf $q$-Virasoro Algebra and the Point-Splitting}\\[20pt]
M. Chaichian\footnote{High Energy Physics Laboratory, Department of Physics
and Research Institute for High Energy Physics, University of Helsinki,
P.O. Box 9 (Siltavuorenpenger 20 C), FIN-00014 Helsinki, Finland}  and
P. Pre\v{s}najder\footnote{Department of Theoretical Physics, Comenius
University, Mlynsk\'{a} dolina, SK-84215, Bratislava, Slovakia}\\[50pt]
{\bf Abstract}
\end{center}
It is shown that a particular $q$-deformation of the
Virasoro algebra can be interpreted in
terms of the $q$-local field $\Phi (x)$  and the
Schwinger-like point-splitted Virasoro currents, quadratic in
$\Phi (x)$. The $q$-deformed Virasoro algebra possesses an additional
index $\alpha$, which is directly related to point-splitting of the
currents. The generators in the $q$-deformed case are found to exactly 
reproduce the results obtained by 
probing the fields $X(z)$ (string coordinate) and $\Phi (z)$
(string momentum) with the non-splitted Virasoro generators and lead to 
a particular representation of the $SU_q (1,1)$ algebra characterized 
by the standard conformal dimension $J$ of the field. Some remarks  
concerning the $q$-vertex operator for the interacting $q$-string theory 
are made.
\end{titlepage}

\vspace{3cm}

\newpage
\section{Introduction}

The conformal  mappings of a  complex plane $\bf C$  belong to a classical
part of mathematics  which has  found  various  applications  in  physics.
One  of  the recent promising applications is  related to  conformal  field
and string theories, leading  to a better  understanding
and to new methods in 2D-quantum field theory (see \cite{GSW}).

The conformal mappings of a complex plane $\bf C$
\be
z \to z' = z\ -\ \sum_{n=0}^{\infty} a_n z^n
\ee
are  generated  in  the neighborhood  of  identity  mapping  by
differential operators
\be
{\cal L}_n \ =\ -z^{n+1} \partial_z \ ,\ n = 0,1,2, \dots \ ,
\ee
acting on  a suitable set of  holomorphic functions $\Phi(z)$,
$z \in {\bf C}$. They form a basis of the Witt algebra and satisfy the
commutation relations
\be
[{\cal L}_n ,{\cal L}_m ]\ =\ (n-m) {\cal L}_{m+m} \ ,\
n, m = 0, 1, 2, \dots \ .
\ee

The  Virasoro   algebra  is  obtained   by  extending  the commutation
relations to  all  $n, m \in {\bf Z}$, followed  by a central extension of the
algebra. Denoting the generators of  the  Virasoro  algebra   by
$L_n$, $n \in {\bf Z}$, their  defining relations are
\be
[ L_n , L_m ]\ =\ (n-m) L_{m+m} \ +\
\frac{c}{12} (n^3 -n) \delta_{n+m,0} \ ,\ n, m = 0,\pm 1,\pm 2, \dots \ .
\ee
where $c$ is the central  element commuting  with all $L_n$. We would like
to stress that in any irreducible  representation $c$ can be  put to be
a constant, which is real in the unitary representation $L^+_n = L_{-n}$.

The  unitary  irreducible   representations  of  the Virasoro algebra are well
known, \cite{Kac}, \cite{FF}. The simplest one is given by the Sugawara
construction of $L_n$ in terms of an infinite set of free bosonic operators
$a_n , a^+_n = a_{-n}$, $n = 1,2, \dots$, satisfying the commutation
relations
\be
[ a_n , a_m ] \ =\ n \delta_{n+m,0}
\ee
(here we have excluded the zero modes inessential for us) . The
Sugawara formula for $L_n$ reads
\be
L_n \ =\ - \frac{1}{2} \sum :a_k a_m : \delta_{k+m,n} \ ,
\ee
where the summation goes over $k,m =\pm 1,\pm 2,\dots$ (and this convention is
used  in  what  follows too).  The Sugawara formula   (6) leads to the unitary
irreducible representation of the Virasoro algebra (4) with $c =1$.

Introducing a primary free field
\be
X(z) \ =\ \sum \frac{i}{n}\ a_n \ z^{-n} \ ,
\ee
one can straightforwardly show that
\be
[ L_n , X(z) ]\ =\ ({\cal L}_n X)(z) \ ,
\ee
where ${\cal L}_n$ is the differential  operator given in (2). Thus, the
adjoint action  of $L_n$  on  the  field $X(z)$ realizes the centreless
Virasoro  algebra in terms of differential operators ${\cal L}_n$, 
$n \in{\bf Z}$, given explicitly in (2). The differential
operators on  r.h.s. of (8) depend  on the
 conformal dimensions  of  the  fields in
question. If the field $A(z)$ has the conformal dimension $J$, then
\be
[ L_m , A(z) ]\ =\ ({\cal L}^J_m A)(z) \ ,
\ee
where (see \cite{GSW})
\be
{\cal L}^J_m \ =\  -z^{m+1} \partial_z \ +\ m J z^m \ .
\ee
This  can  serve  as   the  definition  of  the  conformal dimension $J$. For
example, the field $X(z)$, the string coordinate, has $J = 0$, whereas the
field
\be
\Phi (z) \ =\ \sum \ a_n \ z^{-n} \ ,
\ee
the string momentum, has $J = 1$.  This is an important field, which allows  to
express the Virasoro current (related  to the string energy - momentum) as
\be
\ L(z) \ =\ \sum \ L_m \ z^{-m} \ =\ \frac{1}{2} :\Phi^2 (z):\ .
\ee

{\it Note}: Putting ${\tilde A}(z) = z^{-J} A(z)$, we obtain
\be
[ L_m ,{\tilde A}(z) ]\ =\ ({\tilde {\cal L}}^J_m A)(z) \ ,
\ee
with
\be
{\tilde {\cal L}}^J_m \ =\  -z^{m+1} \partial_z \ +\ (m+1) J z^m \ .
\ee
For $m = 0,\pm 1$, we recognize in (14) the Lie algebra representation  of the
group $SU(1,1)$ (or its covering ${\tilde {SU}}(1,1)$) from the discrete series
with  the Bargmann index $J$. The transformations generated by
${\tilde {\cal L}}^J_m$ (or ${\cal L}^J_m$), $m = 0,\pm 1$, correspond to the
M\"{o}bius subset of conformal mappings.

In mathematics and physics it is of  great importance to  construct and
interpret the ($q$-)deformations of various algebraic structures. In the case
in question there are a few different $q$-deformations of the infinite
algebras introduced above. Let us mention two of them.

The first  one is related to  the $q$-deformation of the Witt algebra (see
\cite{CurZach}-
\cite{Liu1}) and is given by the formula
\be
{\hat {\cal L}}^{CZ}_m \ =\ \frac{z^m}{\omega (q)}\ (D(q^2 )-1)\ ,
\ m \in {\bf Z} \ ,
\ee
where $\omega (q) = q - q^{-1}$ and $D(q)$ is  the dilatation operator acting
on a suitable set of functions as $D(q) \Phi (z) = \Phi (zq)$, $q$ - (real)
parameter. The relation $D(q^{\alpha} ) = D^{\alpha} (q)$ holds. The  set of
operators ${\hat {\cal L}}^{CZ}_m$, $m \in {\bf Z}$, satisfies the
$q$-commutation relations
\be
q^{m-m'} {\hat {\cal L}}^{CZ}_m {\hat {\cal L}}^{CZ}_{m'} \ -\
q^{m'-m} {\hat {\cal L}}^{CZ}_{m'} {\hat {\cal L}}^{CZ}_m \ =\
[m-m'] {\hat {\cal L}}^{CZ}_{m+m'} \ ,
\ee
where $[x] = \omega(q^x )/\omega(q)$. The analogue of their central extension
was found by \cite{AiSa}, \cite{ChIsLuPoPr} and \cite{Liu1}. The
Harish-Chandra modules of the extended algebra were recently classified in
\cite{Liu2},  but the $q$-analogue of Sugawara construction is still missing.

However, the operators ${\hat {\cal L}}^{CZ}_m$, $m \in {\bf Z}$, do not
close under the commutation and an  additional index $\alpha \in {\bf Z}$ is
needed. Putting
\be
{\hat {\cal L}}^{\alpha}_m \ =\ \frac{z^m}{\omega (q)}\
[ q^{\frac{\alpha m}{2}} D(q^{\alpha} )  -
q^{-\frac{\alpha m}{2}} D(q^{-\alpha} ) ]\ ,
\ m, \alpha \in {\bf Z} \ ,
\ee
we obtain the commutation relations
\be
[{\hat {\cal L}}^{\alpha}_m , {\hat {\cal L}}^{\alpha'}_{m'} ]\ =\
\sum_{\sigma'} [\frac{1}{2}(\alpha m' - \sigma' \alpha' m)]
\frac{[\alpha +\sigma' \alpha' ]}{[\alpha ][\sigma' \alpha' ]}
{\hat {\cal L}}^{\alpha +\sigma' \alpha'}_{m+m'} \ ,
\ee
where the summation runs over $\sigma' = \pm 1$ (see \cite{H-OMNSai},
\cite{ChIsLuPoPr}.
 
In \cite{ChPr} we found the central extension of the algebra (18) based on 
the $q$-analogue of Sugawara construction in terms of a system of fermionic 
oscillators $b_r , b^+_r = b_{-r}, r = 1/2 , 3/2, \dots$, satisfying the 
standard anticommutation relations
\be
[ b_r , b_s ]_+ \ =\ \delta_{r+s,0} \ .
\ee
The operators $L^{\alpha}_m$ satisfying the commutation relations (18) are
given as
\be
L^{\alpha}_m \ =\ L^{-\alpha}_m \ =\ \frac{1}{2\omega (q^{\alpha} )}
\sum q^{\alpha (s-r)/2} :b_r b_s :\ \delta_{r+s,m} \ ,
\ee
where the summation is over $r, s = \pm 1/2, \pm 3/2,\dots$.

This allows us to introduce the free fermionic field
\be
\Psi (z)\ =\ \sum b_r z^{-r} \ ,
\ee
and to express the Virasoro currents
\be
L^{\alpha} (z) \ =\ \sum L^{\alpha}_m z^{-m}
\ee
as  bilinear  expressions  in $\Psi (z)$ and  their $q$-deformed derivative.
This approach was used by \cite{Sato}, in order to develop the operator
product expansion (OPE)  techniques for the $q$-deformed  Virasoro algebras.

However, we need a bosonic realization of the $q$-analogue of Sugawara
construction in order to perform a $q$-analogue of the program described in
the first part of the introduction. Such a realization  we proposed in
\cite{ChPr} and it led to a central extension of a slightly more complicated
$q$-Witt algebra as in (18). This is our starting point in the present article.

Using a proper infinite set of bosonic oscillators proposed in \cite{ChPr},
we introduce the $q$-analogues of the fields $X(z)$  and $\Phi (z)$, and we
show that the additional index $\alpha$ corresponds exactly to the
point-splitting of the arguments of fields entering the Virasoro current (12).
The point-splitting approach to singular expressions was proposed  by Dirac
and Peierls \cite{Di}, \cite{Pe} in early days of QFT. Schwinger used it
systematically in order to regularize the bilinear expressions in fields
(e.g. the currents) in QFT, \cite{Schw}. Calculations based on 
point-splitting were performed in QED 
(see \cite{Schw}, \cite{BouJa}, \cite{Ch}) and
recently in the framework of standard model (see \cite{OW} and \cite{NOW}).
 
In Sect.2 we demonstrate how the $q$-deformation of the set of bosonic
oscillators leads to the point-splitting of Virasoro currents. In Sect.3 we
introduce the notion of the ($q$-)conformal dimensions into our scheme.
We show that the basic features of the proposal in \cite{BeLec} concerning
the $SU_q (1,1)$ deformation of the $SU(1,1)$ symmetry	of M\"{o}bius
mappings actually take place. Finally, Sect.4 contains the concluding remarks.

\section{$q$-deformed Sugawara construction and current point-splitting}

We start, like in \cite{ChPr}, from the infinite set of bosonic operators
$a_n , a^+_n = a_{-n}$, $n = 1,2, \dots$, satisfying the commutation
relations
\be
[ a_n , a_m ] \ =\ [n] \delta_{n+m,0} \ .
\ee
>From this system of oscillators  we construct the auxiliary fields, string
coordinate and string momentum by
\be
{\tilde X}(z) \ =\ \sum \frac{i}{[n]}\ a_n \ z^{-n} \ =\ X(z) \ ,
\ee
and
\be
{\tilde \Phi}(z) \ =\ \sum\ a_n \ z^{-n} \ =\ z \Phi (z) \ .
\ee

The particular $n$-dependent factor on r.h.s. of (23) and in (24) has  the
consequence that ${\tilde X}$ and ${\tilde \Phi}$ fields do not satisfy
the standard commutation relations but the deformed ones:
\be
[ {\tilde X}(z) ,{\tilde X}(w) ]\ =\ \sum \ \frac{i}{[n]}\
(\frac{w}{z} )^n \ ,
\ee
\be
[ {\tilde X}(z) ,{\tilde \Phi}(w) ]\ =\ i \delta (z-w)\ ,
\ee
\be
[ {\tilde \Phi}(z) ,{\tilde \Phi}(w) ]\ =\ \frac{1}{\omega (q)}\
[ \delta (w-zq^{-1} )\ - \delta (w-zq) ]\ ,
\ee
where we have introduced the $\delta$-function by
\be
\delta (z-w)\ =\ \frac{1}{w} \sum \ (\frac{w}{z} )^n \ =\
\frac{1}{z} \sum \ (\frac{z}{w} )^n \ .
\ee
We stress that	the r.h.s. of (26) can	be expressed as an infinite series
of $\delta$-functions: $\delta (w-zq^{2k+1} )$, $k \in {\bf Z}$. If we adopt
the terminology that a field is {\it q-local} when the r.h.s. of its
commutator contains a finite number of terms $\delta (w-zq^n )$,
$n \in {\bf Z}$, then we see that the  field ${\tilde \Phi}(z)$ is  local,
whereas ${\tilde X}(z)$ is not.

{\it Note}: This is consistent with the standard string theory: for
$q\to 1$, the r.h.s. of (28)  approaches ${\delta}'(z-w)$, whereas the r.h.s.
of (26) in this limit has the sign function $\varepsilon (z-w)$.

In \cite{ChPr} we introduced a bosonic realization of the $q$-Sugawara
generators by defining
\be
L^{\alpha}_m \ =\ L^{-\alpha}_m \ =\ \frac{1}{2}
\sum q^{\alpha (n-k)/2} :a_k a_n :\ \delta_{k+n,m} \ ,
\ee
and we showed that they satisfy the commutation relations
\be
[L^{\alpha}_m , L^{\alpha'}_{m'} ]\ =\ \frac{1}{4}\sum_{\sigma ,\sigma'}
[\frac{1}{2} (m-m'-m'\sigma \alpha +m\sigma' \alpha']\
L^{\sigma \alpha +\sigma' \alpha' +1}_{m+m'} \ +\ C.T.\ ,
\ee
where the symbol $C.T.$ denotes the central term which is inessential for our
purposes (it can be found in \cite{ChPr}). The summation in (31) runs over
$\sigma ,\sigma' = \pm 1$ (i.e., it contains four terms compared with
two in the fermionic realization (18)). It is important to mention  that
the introduction of the index $\alpha$ is inevitable. For
$\alpha \in {\bf Z}$ the algebra (31) closes (the minimal set for which (31)
closes is formed  by $\alpha$ - odd integer).

Introducing now the Virasoro currents
\be
L^{\alpha} (z) \ =\ \sum L^{\alpha}_m z^{-m} \ =\
z^2 {\tilde L^{\alpha} (z)} \ ,
\ee
one can straightforwardly derive the $q$-analogue of (12):
\be
L^{\alpha} (z) \ =\
\frac{1}{2} :\Phi (zq^{\alpha /2} ) \Phi (zq^{-\alpha /2} ) :\ ,
\ee
\be
{\tilde L}^{\alpha} (z) \ =\ \frac{1}{2}
:{\tilde \Phi}(zq^{\alpha /2} ) {\tilde \Phi}(zq^{-\alpha /2} ) :\ .
\ee
We see that the additional index $\alpha$ directly measures the splitting
of arguments of the fields entering the Virasoro currents. Recalling that
the unit circle $|z| = 1$ is related to the light-cone local string world sheet
parameters, the point-splitting (33) (or (34)) can be interpreted as the
Schwinger point-splitting \cite{Sch} of the currents in complex direction in
the complexified 2D space-time.

using the commutation relations (28), one can easily show that all fields
of the form
\be
:\Phi (zq^{\alpha_1} )\ \dots \ \Phi (zq^{\alpha_N} ) :\ ,\
N = 1,2,\dots \ ,
\ee
are  mutually $q$-local, i.e. their commutators contain finite linear
combinations of terms containing the $\delta$-functions $\delta (w-zq^n)$,
$n \in {\bf Z}$. In particular,
\[
[{\tilde L}^{\alpha} (z) ,{\tilde L}^{\alpha} (w) ]\ =\
\frac{1}{2\omega (q)z} \sum_{\sigma ,\sigma'} \
[\delta (wq^{-\sigma' \alpha' /2} -zq^{1-\sigma \alpha /2})
{\tilde L}^{\sigma \alpha +\sigma' \alpha' +1} (zq^{(1+\sigma' \alpha')/2} )
\]
\be
-\ \delta (wq^{-\sigma' \alpha' /2} -zq^{-1-\sigma \alpha /2})
{\tilde L}^{\sigma \alpha +\sigma' \alpha' +1} (zq^{-(1+\sigma' \alpha')/2}
\ +\ C.T.
\ee
and this is in full agreement with (31).

Instead of  using commutators, the OPE formalism can be  developed along
similar lines as in the  usual nondeformed case ($q = 1$). We shall not pursue
this point but instead  we shall introduce in the next section the notion
of conformal dimension into our model.

{\it Note 1}:  There are different approaches motivated by $q$-affine
algebras \cite{D},\cite{FJ}, \cite{Be}, in  which the canonical realizations
contain an infinite set of oscillators satisfying commutation relations
similar to (23) but with more complicated functions on r.h.s. This  can lead
to a more complicated regularization of the current (terms proportional to
$q^{\pm m}$ are responsible for point splitting, factors containing $n$ for
derivatives, etc.). Our choice is very simple, and this is the  reason why it
leads to the pure point-splitting.

{\it Note 2}: We would like to stress that the Virasoro currents (22)
investigated in \cite{Sato} are expressed in terms of a local fermionic field
${\tilde \Psi} (z)$,
\be
[ {\tilde \Psi}(z) ,{\tilde \Psi}(w) ]_+ \ =\ \sum (\frac{w}{z} )^n \ =\
w \delta (z-w)\ ,
\ee
but they themselves are expressed in the splitted form
\be
{\tilde L}^{\alpha} (z) \ =\ - \frac{1}{2\omega (q^{\alpha})}
:{\tilde \Psi}(zq^{\alpha /2} ) {\tilde \Psi}(zq^{-\alpha /2} ) :\ .
\ee
If  instead of (19) one starts from the fermionic oscillators satisfying
\be
[ b_r , b_s ]_+ \ =\ \frac{1}{2} (q^r + q^{-r} ) \delta_{r+s,0} \ ,
\ee
one arrives at the $q$-local fermionic field satisfying
\[
[ {\tilde \Psi}(z) ,{\tilde \Psi}(w) ]_+ \ =\
\frac {1}{2} \sum \ (\frac{wq}{z} )^n \ +\ \frac {1}{2} \sum \ (\frac{w}{qz} )^n
\]
\be
=\ \frac{w}{2} [\delta (w-zq^{-1} )\ +\ \delta (w-zq)] \ .
\ee
However, the formula (38) for the splitted currents remains unchanged. Of
course, the $q$-Virasoro algebra becomes more complicated, and up to  sign
corrections on r.h.s., it is identical to (36).

\section{Splitted Virasoro algebra and $q$-conformal dimensions}

The notion of generalized q-conformal Ward identities was introduced and
investigated in \cite{BeLec}. One of the basic points has been the extension
to the $q$-deformed case of the notion of conformal weight $J$, defined in the
non-deformed case either by
\[
[ L_m , A(z) ]\ =\ ({\cal L}^J_m A)(z) \ ,
\]
\be
{\cal L}^J_m A(z)\ =\ [-z^{m+1} \partial_z \ +\ m J z^m ]A(z)\ ,
\ee
or for ${\tilde A}(z) = z^{-J} A(z)$ by
\[
[ L_m ,{\tilde A}(z) ]\ =\ ({\tilde {\cal L}}^J_m A)(z) \ ,
\]
\be
{\tilde {\cal L}}^J_m {\tilde A} (z)\ =\
[-z^{m+1} \partial_z \ +\ (m+1) J z^m ]{\tilde A} (z)\ .
\ee
The subset of operators ${\tilde {\cal L}}^J_m$, $m = 0,\pm 1$, just
corresponds to the irreducible representation of SU(1,1) group (from a
discrete series of unitary representations with the Bargmann index $J$).

The idea in \cite{BeLec} was to use instead of this representation the
$q$-deformed analogue given as
\[
{\tilde {\cal L}}^J_{+1} {\tilde A} (z)\ =\ \frac{z}{\omega (q)}
[q^{2J} D(q) - q^{-2J} D(q^{-1} )]{\tilde A} (z)\ ,
\]
\[
{\tilde {\cal L}}^J_{-1} {\tilde A} (z)\ =\ \frac{z^{-1}}{\omega (q)}
[ D(q) - D(q^{-1} )]{\tilde A} (z) \ ,
\]
\be
{\tilde {\cal K}}^J_0 {\tilde A} (z)\ =\ q^J D(q) {\tilde A} (z) \ .
\ee
These operators satisfy the $SU_q (1,1)$ - algebra relations
\[
{\tilde {\cal K}}^J_0 {\tilde {\cal L}}^J_{\pm 1} \ =\ q^{\pm 1}
{\tilde {\cal L}}^J_{\pm 1} {\tilde {\cal K}}^J_0 \ ,
\]
\be
[ {\tilde {\cal L}}^J_{+1} ,{\tilde {\cal L}}^J_{-1} ]\ =\ -\frac{1}{\omega(q)}\
[({\tilde {\cal K}}^J_0 )^2 \ -\ ({\tilde {\cal K}}^J_0 )^{-2} ]\ .
\ee
Equivalently, one can use the fields $A(z)$. Then the operators ${\cal K}^J_0$,
${\cal L}^J_{\pm 1}$, satisfying (44) are given as follows :
\[
{\cal K}^J_0 A(z)\ =\ D(q) A(z) \ ,
\]
\be
{\cal L}^J_m A(z)\ =\ z^m [{\cal D}^J_m (q) - {\cal D}^J_m (q^{-1} )]A(z)\ ,
\ee
where
\be
{\cal D}^J_m (q) \ =\ \frac{q^{mJ}}{\omega (q)}\ D(q)\ .
\ee

{\it Note}: The restriction to the M\"{o}bius subgroup of mappings made in
\cite{BeLec} was caused by the fact that the $SU(1,1)$ algebra has the well
defined deformation (as a Hopf algebra) to $SU_q(1,1)$, whereas for the
Virasoro algebra such a deformation is not known.

In our model we have at our disposal the set of generators $L^{\alpha}_m$. We
can calculate the commutators $[L^{\alpha}_m , A(z)]$ for some particular
chosen $A(z)$ to see whether eqs. (45) are satisfied or in what way they are
modified.

Let us take first for $A(z)$ the elementary fields $X(z)$ and $\Phi (z)$.
Using eqs. (23) and (30) we obtain
\[
[L^{\alpha}_m , X(z)] \ =\ z^m \ \sum_{\sigma ,\sigma'}
\frac{q^{-\sigma \sigma' \alpha m/2}}{2\omega (q^{\sigma})}\
X(zq^{-\alpha \sigma \sigma'+\sigma})
\]
\be
=\ z^m \ \{ \frac{q^{-\alpha m/2}}{2\omega (q)}\ X(zq^{-\alpha +1})\ +\
\frac{q^{\alpha m/2}}{2\omega (q)}\ X(zq^{\alpha +1})\ +\
(q \to q^{-1} ) \} \ ,
\ee
where the  summation runs over $\sigma ,\sigma' = \pm 1$, and the
$(q \to q^{-1} )$ denote terms written down inside the curly brackets but
with $q$ replaced by $q^{-1}$. Similarly,
\[
[L^{\alpha}_m , \Phi(z)] \ =\ z^m \ \sum_{\sigma ,\sigma'}
\frac{q^{\sigma m-\sigma \sigma' \alpha m/2}}{2\omega (q^{\sigma})}\
\Phi (zq^{-\alpha \sigma \sigma'+\sigma})
\]
\be
=\ z^m \ \{ \frac{q^{m-\alpha m/2}}{2\omega (q)}\ \Phi(zq^{-\alpha +1})\ +\
\frac{q^{m+\alpha m/2}}{2\omega (q)}\ \Phi(zq^{\alpha +1})\ +\
(q \to q^{-1} ) \} \ .
\ee
We see  that the action of $L^{\alpha}_m$ on the fields $A(z) = X(z)$ and
$\Phi (z)$ can be realized in terms of infinite matrices over the space
\[
V_1 (A)\ =\ Span\{ A_{\beta} (z,q)\ =\ A(zq^{\beta} )\ ,\ \beta \in {\bf Z}\} \ 
,
\]
where $A(z) = X(z)$ or $\Phi (z)$.

Using the dilatation operator $D(q)$, we can rewrite eqs. (47) and (48) in the
form
\[
[L^{\alpha}_m , X(z)] \ =\ z^m \ \sum_{\sigma ,\sigma'}
\frac{q^{-\sigma \sigma' \alpha m/2}}{2\omega (q^{\sigma})}\
D(q^{-\alpha \sigma \sigma'+\sigma}) X(z)
\]
\be
=\ z^m {\cal D}^0_m (q)\ \frac{1}{2}\{ q^{-\alpha m/2} D(q^{-\alpha})\  +\
q^{\alpha m/2} D(q^{\alpha})\ \} X(z)\ +\ (q \to q^{-1} ) \ ,
\ee
and
\[
[L^{\alpha}_m , \Phi (z)] \ =\ z^m \ \sum_{\sigma ,\sigma'}
\frac{q^{\sigma m-\sigma \sigma' \alpha m/2}}{2\omega (q^{\sigma})}\
D(q^{-\alpha \sigma \sigma'+\sigma}) \Phi (z)
\]
\be
=\ z^m {\cal D}^1_m (q)\ \frac{1}{2}\{ q^{-\alpha m/2} D(q^{-\alpha})\  +\
q^{\alpha m/2} D(q^{\alpha})\ \} \Phi (z)\ +\ (q \to q^{-1} ) \ .
\ee
Comparing eqs.(49) and (50) with eqs.(45) we see that for $\alpha = 0$ we
reproduce the result of \cite{BeLec} for $J = 0$ and $J = 1$, respectively. For
$\alpha \neq 0$ we have an $\alpha$-dependent correction as a result of the
fact that we probe the field in question with $L^{\alpha}_m$ corresponding to
the splitted Virasoro currents. It is therefore reasonable to assign
 to the
fields $X(z)$ and $\Phi (z)$ the $q$-conformal dimensions $J=0$ and $J=1$,
respectively.

As the next interesting possibility we shall investigate quadratic fields of the
form $\Phi^2_{\beta_1 \beta_2} (z;q)=\Phi (zq^{\beta_1}) \Phi (zq^{\beta_2})$,
which span the space $V_2 (\Phi ) = V_1 (\Phi ) \otimes V_1 (\Phi )$. 
We denote $V^0_2 (\Phi )$ as the subspace spanned by
$\Phi^2_{\beta_1 \beta_2} (z;q)$ with $\beta_1 + \beta_2 = 0$.

Using the relation
\[
[L^{\alpha}_m , \Phi (zq^{\beta_1}) \Phi (zq^{\beta_2})]\
\]
\[
=\ [L^{\alpha}_m , \Phi (zq^{\beta_1}) ] \Phi (zq^{\beta_2})\ +\
\Phi (zq^{\beta_1}) [L^{\alpha}_m , \Phi (zq^{\beta_2})]\ ,
\]
and eqs.(48), we obtain
\[
[L^{\alpha}_m , \Phi^2_{\beta_1 \beta_2} (z;q)]\
\]
\[
=\ z^m \ \{ \frac{q^{m-\alpha m/2}}{2\omega (q)}\
\{ q^{m\beta_1} \Phi^2_{\beta_1 -\alpha +1,\beta_2} (z;q) \ +\
q^{m\beta_2} \Phi^2_{\beta_1 ,\beta_2 -\alpha +1,} (z;q) \} \
\]
\be
+\ z^m \ \{ \frac{q^{m+\alpha m/2}}{2\omega (q)}\
\{ q^{m\beta_1} \Phi^2_{\beta_1 +\alpha +1,\beta_2} (z;q) \ +\
q^{m\beta_2} \Phi^2_{\beta_1 ,\beta_2 +\alpha +1,} (z;q) \} \ +\
(q \to q^{-1} ) \} \ .
\ee
This equation defines the action of $L^{\alpha}_m$ on
$\Phi^2_{\beta_1 \beta_2} (z;q)$ as an infinite matrix on the space
$V_1 (\Phi ) \otimes V_1 (\Phi )$.

In order to assign the $q$-conformal dimensions of quadratic fields
$\Phi^2_{\beta_1 \beta_2} (z;q)$, we restrict  ourselves to fields from
$V^0_2 (\Phi )$ centered around $z$, and we rewrite the r.h.s. of (51) in
the form of some difference operators acting on the centered quadratic fields:
\[
[L^{\alpha}_m , \Phi^2_{\beta_1 \beta_2} (z;q)]\
\]
\[
=\ z^m  \{ \frac{q^{m-\alpha m/2}}{2\omega (q)}\ D(q^{(1-\alpha)/2} )
\{ \Phi^2_{\beta_1 +(1-\alpha )/2,\beta_2 -(1-\alpha )/2} (z;q)
\]
\[
+\ \Phi^2_{\beta_1 -(1-\alpha )/2,\beta_2 +(1-\alpha )/2,} (z;q) \} \
\]
\[
+\ z^m \ \{ \frac{q^{m+\alpha m/2}}{2\omega (q)}\ D(q^{(1+\alpha)/2} )
\{ \Phi^2_{\beta_1 +(1+\alpha )/2,\beta_2 -(1+\alpha )/2} (z;q)
\]
\be
+\ \Phi^2_{\beta_1 -(1+\alpha )/2,\beta_2 +(1+\alpha )/2,} (z;q) \} \ +\
(q \to q^{-1} ) \} \ .
\ee
This expression can be put into the required form by using $q^{1/2}$ as the
deformation parameter  instead of $q$. Using
$\Phi^2_{\beta_1 \beta_2} (z;q) = \Phi^2_{2\beta_1 ,2\beta_2} (z;q^{1/2} )$
we have:
\[
[L^{\alpha}_m , \Phi^2_{2\beta_1 ,2\beta_2} (z;q^{1/2} )]\ \]
\[
=\ z^m {\cal D}^2_m (q^{1/2} ) \{ q^{-\alpha m/2} D(q^{-\alpha /2} )
R^{2,\alpha}_m \Phi^2_{2\beta_1 ,2\beta_2} (z;q^{1/2} ) \ \]
\be
+\ q^{\alpha m/2} D(q^{\alpha /2} ) R^{2,-\alpha}_m
\Phi^2_{2\beta_1 ,2\beta_2} (z;q^{1/2} ) \} \ +\ (q \to q^{-1} ) \ ,
\ee
where $R^{2,\alpha}_m$ is an operator in $V^0_2 (\Phi )$ defined as
\[
R^{2,\alpha}_m \Phi^2_{2\beta_1 ,2\beta_2} (z;q^{1/2} ) \ =\
\frac{q^{m\beta_1}}{q^{1/2} +q^{-1/2}}
\Phi^2_{2\beta_1 +1-\alpha ,2\beta_2 -1+\alpha } (z;q^{1/2} ) \
\]
\be
+\ \frac{q^{m\beta_2}}{q^{1/2} +q^{-1/2}}
\Phi^2_{2\beta_1 -1+\alpha ,2\beta_2 +1-\alpha } (z;q^{1/2} ) \ .
\ee

We see that it is reasonable to assign to the field
$\Phi^2_{\beta_1 \beta_2} (z;q)$ the $q$-conformal dimensions $J = 2$.
The analogous formula is valid for the field $X^2_{\beta_1 \beta_2} (z;q)$
with the only change of ${\cal D}^0_m (q^{1/2} )$ instead of
${\cal D}^2_m (q^{1/2} )$ in (53). Therefore, we assign to the field
$X^2_{\beta_1 \beta_2} (z;q)$ the conformal dimension $J = 0$. These
values for the $q$-conformal dimensions have been expected, since they are just
the same as those in the standard case. However, what was perhaps not
expected, is the redefinition of the deformation parameter from $q$ to
$q^{1/2}$.

{\it Note}: The procedure described above can be directly generalized to
the higher field polynomials. The resulting formula for
$\Phi^N_{\beta_1 \dots \beta_N} (z;q)
=\Phi^N_{N\beta_1 ,\dots ,N\beta_N} (z;q^{1/N} )$ reads:
\[
[L^{\alpha}_m , \Phi^N_{N\beta_1 ,\dots ,N\beta_N} (z;q^{1/N} )]\
\]
\[
=\ z^m {\cal D}^N_m (q^{1/N} ) \{ q^{-\alpha m/2} D(q^{-\alpha /N} )
R^{N,\alpha}_m \Phi^N_{N\beta_1 ,\dots ,N\beta_N} (z;q^{1/N} ) \
\]
\be
+\ q^{\alpha m/2} D(q^{\alpha /N} ) R^{N,-\alpha}_m
\Phi^N_{N\beta_1 ,\dots ,N\beta_N} (z;q^{1/N} ) \} \ +\ (q \to q^{-1} ) \ ,
\ee
where $R^{N,\alpha}_m$ is a linear  operator in the space $V^0_N (\Phi )$
with a similar (but slightly more complicated) expression as
$R^{2,\alpha}_m$ given in (54). The same formula is valid for
$X^N_{\beta_1 \dots \beta_N} (z;q)$ but with ${\cal D}^0_m (q^{1/N} )$
instead of ${\cal D}^N_m (q^{1/N} )$ in (55). Therefore, we assign to
$\Phi^N_{\beta_1 \dots \beta_N} (z;q)$ the $q$-conformal dimension $J = N$,
whereas for $X^N_{\beta_1 \dots \beta_N} (z;q)$, we put $J = 0$. However,
we should recall that the deformation parameter in both cases  is
$q^{1/N}$, and not $q$. This result can be easily understood: the action
of the operator $L^{\alpha}_m$ modifies successively the factors
$q^{\beta}$ in one of the factors in the monomial
$\Phi (zq^{\beta_1} ) \dots \Phi (zq^{\beta_N} )$ to
$q^{\beta \pm 1\pm \alpha}$. This indeed means that the mean argument
$z = (\prod zq^{\beta_j} )^{1/N}$ of  the centered monomial
$\Phi (zq^{\beta_1} ) \dots \Phi (zq^{\beta_N} )$ is changed by the  factors
$q^{(\pm 1\pm \alpha )/N} )$, and this leads to the effective change of the
deformation parameter $q$ to $q^{1/N}$ in (55).

It is well known that the powers $X^N (z)$, and even analytic functions
$F(X(z))$, all have in the usual nondeformed case conformal dimension
$J =0$. However the normal ordering spoils this direct dimension counting and
$:F(X(z)):$ may acquire non-zero conformal dimensions (see e.g. \cite{GSW}).
The best known example is the (tachyon) vertex function
\be
V(z,k)\ =\ :\exp (ikX(z)): \ ,
\ee
which has the conformal dimension $J =\frac{1}{2} k^2$.

Of course, the normal ordering modifies formulas like (55) determining the
conformal dimension (of powers of primary fields  $X(z)$ or $\Phi (z)$. It
would be of much interest to find $q$-deformed analogue of the vertex operator
(56) possessing non-zero $q$-conformal dimension. This interesting and
well-defined problem is under study.

\section{Comments and conclusions}

In this paper we first show that the particular $q$-deformation of the
Virasoro algebra introduced previously  in \cite{ChPr} can be interpreted in
terms of the $q$-local fields (either bosonic, or fermionic) and the splitted
Virasoro currents, which are quadratic in the fields. Our $q$-deformed Virasoro
algebra possesses one additional index, which is directly related to the
Schwinger-like point-splitting of the arguments of currents. We have
demonstrated how this works for the bosonic
realization. As well it works also for the fermionic realization proposed
in \cite{ChPr} and further used in \cite{Sato}. Actually, similar
point-splitting would appear in any  free field realization (Sugawara
construction) of currents if in the expressions for the latter appear
factors such as $q^{\pm m}$ (additional powers of $m$ lead to usual
derivatives of fields). It would be interesting to investigate from this
point of view the affine Kac-Moody algebras free field realizations
\cite{D}, \cite{FJ}, \cite{Be}.

In this paper we have restricted ourselves to all the constructions with a real
deformation parameter $q$. However, there exists a different branch of
realizations of various $q$-deformed structures corresponding to the choice
of $q$ being root of unity, $q = e^{2\pi i/N}$ (see e.g. \cite{Hay},
\cite{A-GGS}). In our case this would lead to a particular $q$-deformation
of the Virasoro algebra containing only a finite set of generators
$L^{\alpha}_m$, $\alpha ,m$ both integer and $|\alpha |, |m| \le N$. We have
not yet investigated this interesting case in detail.

Naturally, there appears the possibility to extend the point-splitting
approach to Virasoro superalgebras or to Kac-Moody algebras. We have
investigated the superalgebras related to $q$-deformed (equivalently,
splitted) Virasoro algebra. The superalgebras we have found have the
following structure: let $L'$ and $L''$ denote the bosonic and fermionic
realization of even generators respectively, and let G be the set of odd
generators of Virasoro $q$-superalgebra. Then, we obtain that
\[
[L' ,L' ] \subset L' \ ,\ [L'' ,L'' ] \subset L'' \ ,
\]
\[
[L' ,G ] \subset G \ ,\ [L'' ,G ] \subset G \ ,
\]
\[
[G ,G ] \subset L' + L'' \ ,\ [L' ,L'' ] = 0 \ .
\]
However, the structure constants in the bosonic and fermionic realizations
appearing in the r.h.s. of the above equations are different from each other
in contrast to the usual case of Virasoro superalgebra. Thus, one can not
introduce a single algebra $L = L' + L''$ as in the non-deformed case. The
usual (nondeformed) Virasoro superalgebra generated by even generators
$L$ and odd ones $G$ has the standard structure
\[
[L, L] \subset L \ ,\ [L, G] = G \ ,\ [G, G] \subset L \ .
\]
In our scheme this property can not be recovered (except for the $q \to 1$
limit).

From the point of view of the splitted Virasoro algebra, this unexpected
result can be understood as follows. The point-splitting introduces a
particular regularization into the theory, and this regularization forbids
the standard supersymmetric  extension. Similar features are shared by
lattice regularization: it is well known that the supersymmetry can not
be implemented on a lattice.

The commutators $[L_m , A(z) ] = {\cal L}^J_m A(z)$, $m=0,\pm $1, in the
nondeformed case are directly related to particular representations of the
algebra $SU(1,1)$ characterized just by the conformal dimension $J$ of the
field in question. In \cite{BeLec} it was proposed to use in the
$q$-deformed case the representations of the $q$-deformed algebra
$SU_q (1,1)$ instead of the corresponding ones of the $SU(1,1)$ algebra.
We have shown that in our $q$-deformed model this is indeed the case: for
the primary fields $X(z)$ and $\Phi (z)$ we exactly reproduce the
results of \cite{BeLec} when probing the fields with the non-splitted
Virasoro generator $L^0_m$ (the splitted generators $L^{\alpha}_m$,
$\alpha$-splitting leads to $\alpha$-dependent corrections). The situation
is technically more complicated for the splitted products of primary  fields
$\Phi (zq^{\beta_1} )\dots \Phi (zq^{\beta_N} )$ and
$X(zq^{\beta_1} )\dots X(zq^{\beta_N} )$. However, even in this case the
corresponding formulas contain all the operators of \cite{BeLec} related to
the representations of $SU_q (1,1)$ with $J = N$ and $J = 0$, respectively.
Thus we reproduce the expected result concerning the conformal dimension of
$N$-fold products, with the only difference that the deformation parameter
should be $q^{1/N}$ instead of $q$.

The problem of constructing the $q$-vertex operator $V_q (k,X)$, i.e. the
analogue of $V(k,X) = :exp.(ikX(z)):$, still remains open. In the nondeformed
case  this function, due to normal ordering, has the conformal dimension
$J = \frac{1}{2} k^2$ (despite the fact that $X(z)$ has zero conformal
dimension).

The adjoint action, $[L^{\alpha}_m ,A(z)]={\hat {\cal L}}^{J\alpha}_m A(z)$,
leads to various representations (depending on J) of the centreless version
of the splitted Virasoro algebra. It would be important to find the
co-algebraic structure of the $q$-deformed Virasoro algebra in order to
understand better how various representations are related to each other. In
\cite{ChPr} we have found a non-trivial but co-commutative coproduct for the
$q$-deformed Virasoro algebra. A similar co-commutative coproduct was
proposed in \cite{Saito} but for the centreless Virasoro algebra (18). The
problem  of finding a proper non-cocommutative coproduct still remains
unsolved.

Many formulas in our model become more transparent if one uses  directly the
splitted currents
$L^{\alpha} (z) =\frac{1}{2} :\Phi (zq^{\alpha /2} \Phi (zq^{-\alpha /2}:$.
As a rule , all commutators of $L^{\alpha} (z)$ with splitted polynomials
(ordered or nonordered) in the primary fields contain as
coefficient functions on r.h.s. only the shifted $\delta$-functions
$\delta (z-wq^{\beta} )$. Alternatively, the corresponding OPE in the
coefficient functions contain only the shifted poles $(z-wq^{\beta} )^{-1}$
of the first order, which in the limit $q \to 1$ collapse to non-shifted
higher	order poles $(z-w)^{-n}$.

At  present stage, we are able to formulate a free string model with the
fields $X(z)$ and $\Phi (z)$ as the $q$-string coordinate and the
$q$-string momentum, respectively. However, to introduce interaction, we
need to construct the $q$-vertex operator. Yet, this is one  of the unsolved
problems. Nevertheless, we believe that our  present attempt and similar
models as presented here can shed light on the link between $q$-deformations
and splitting and provide with a basis for regularization in QFT in general.
\newline

{\bf Acknowledgements}
We are grateful to Alan Carey, Katri Huitu and Ruibin Zhang for useful
discussions.
 

\newpage
 \begin{thebibliography}{99}
 \bibitem{GSW} M. B. Green, J. H. Schwarz and E. Witten, Superstring
	      theory  (Cambridge U. P., Cambridge, 1987).
 \bibitem{Kac} V. G. Kac, Lect. Notes in Phys. 94 (1979) 441.
 \bibitem{FF} B. L. Feigin and D. B. Fuchs, Func. Anal. App. 16 (1982) 114.
 \bibitem{CurZach} T. L. Curtright and C. K. Zachos, Phys.Lett. B 243
                  (1990) 237.
 \bibitem{ChKuLu} M. Chaichian, P. P. Kulish and J. Lukierski, Phys.
		   Lett. B 243 (1990) 401.
 \bibitem{Pol} A. P. Polychronakos, Phys. Lett. B 256 (1991) 35.
 \bibitem{AiSa} N. Aizawa and H. Sato, Phys. Lett. B 256 (1991) 185.
 \bibitem{ChIsLuPoPr} M. Chaichian, A. P. Isaev, J. Lukierski,
 Z. Popowicz and P. Pre\v{s}najder, Phys. Lett. B 262 (1991) 32.
 \bibitem{Liu1} K. Q. Liu, C. R. Math. Rep. Ac. Sci. Canad. XIII (1991) 135.
 \bibitem{Liu2} K. Q. Liu, Jour. of Alg. 171 (1995) 606.
 \bibitem{H-OMNSai} H. Hiro-oka, O. Matsui, T. Naito and S. Saito,
                      preprint TMUP-HEL-9004 (1990).
 \bibitem{ChPr} M. Chaichian and P. Pre\v{s}najder, Phys. Lett. B 277 (1992)
               109.
 \bibitem{Sato} H. Sato, Nucl. Phys. B393 (1993) 442.
 \bibitem{Sch} J. Schwinger, Phys. Rev. Lett. 3 (1959) 296.
\bibitem{BeLec} D. Bernard and  A. Leclair, Phys. Lett. B 227 (1989)
                 417.
\bibitem{D} V. G. Drinfeld, Sov. Math. Dokl. 36 (1988) 212.
\bibitem{FJ} I. B. Frenkel and  N. Jing, Proc. Nat. Acad. Sci. USA
	      85 (1988) 9373.
\bibitem{Be}  D. Bernard, Lett. Math. Phys. 17 (1989).
\bibitem{Saito} S. Saito, {\it $q$-Virasoro and $q$-Strings} in:
	 Quarks, symmetries and strings, (World Sci., Singapore, 1991).
%
 \bibitem{Hay} T. Hayashi, Comm. Math. Phys. 127 (1990) 129.
 \bibitem{A-GGS} L.Alvarez-Gaum\'{e}, C. Gomez and G. Sierra, Phys. Lett.
		 B  220 (1989) 142,  Nucl.Phys. B  319 (1989) 155,
                 and B 330 (1990) 347.
 %
  \bibitem{Di} P. A. Dirac, Proc. Cambridge Phil. Soc.30 (1934) 150.
  \bibitem{Pe} R. Peierls, Proc. Roy. Soc., Ser. A 146 (1934) 420.
  \bibitem{Schw} J. Schwinger, Phys. Rev. 82 (1951) 664 and Phys.
		 Rev. Lett. 3 (1959) 296.
  \bibitem{BouJa} D. G. Boulware and R. Jackiw, Phys. Rev. 186 (1969)
		 1442.
  \bibitem{Ch} M. S. Chanowitz, Phys. Rev. D 2 (1970) 3016.
  \bibitem{OW} P. Osland and T. T. Wu, Z. Phys. C 55 (1992) 569 and 585
             and 593.
  \bibitem{NOW} C. Newton, P. Osland and T. T. Wu, Z. Phys. C 61 (1994)
               441.
\end{thebibliography}
\end{document}